\newcommand{\subparagraph}{}

\documentclass[journal]{IEEEtran}
\input{alphabet}   

\usepackage[utf8x]{inputenc}
\usepackage[T1]{fontenc}
\usepackage{verbatim}
\usepackage{lmodern}
\usepackage{graphicx,amssymb,lineno}
\usepackage{amsmath}
\usepackage{graphics}
\usepackage{colortbl}
\usepackage{bbold}
\usepackage{multirow}
\usepackage{slashbox}
\usepackage{rotating}
\usepackage{subfigure}
\usepackage{lipsum}
\usepackage[english]{babel}
\usepackage{color}

\usepackage{titlesec}

\graphicspath{{./Images/}}
\DeclareGraphicsExtensions{.pdf,.png}

\usepackage{url}
\newcommand{\kT}{$k_T$}
\newcommand{\diag}{\mathop{diag}}

\newcommand{\R}{\ensuremath{\mathbb{R}}}
\newcommand{\C}{\ensuremath{\mathbb{C}}}

\newcommand{\mmin}[2]{\mathop{\begin{array}[t]{l} \min \ #1 \\
                                  \begin{array}[t]{l} #2 \end{array} \end{array}}}

\newcommand{\Id}{\mathbf{Id}}
\newcommand{\ds}{\displaystyle}

\titlespacing*{\paragraph}
{0pt}{3.25ex plus 1ex minus .2ex}{1.5ex plus .2ex}
\titlespacing*{\subsubsection}
{0pt}{3.25ex plus 1ex minus .2ex}{1.5ex plus .2ex}

\DeclareSymbolFont{bbold}{U}{bbold}{m}{n}
\DeclareSymbolFontAlphabet{\mathbbold}{bbold}

\title{On Variant Strategies To Solve The Magnitude Least Squares Optimization Problem In Parallel Transmission Pulse Design And Under Strict SAR And Power Constraints}
\author{A. Hoyos-Idrobo, P. Weiss, A. Massire, A. Amadon, N. Boulant %

\thanks{A. Hoyos-Idrobo, A. Massire, A. Amadon and N. Boulant are with CEA, I2BM, NeuroSpin, UNIRS, Gif sur Yvette, France
        {\tt \small nicolas.boulant@cea.fr}}%
\thanks{P. Weiss is with ITAV- USR3505, Universit\'e de Toulouse, CNRS, France
        {\tt\small pierre.armand.weiss@gmail.com}}

\thanks{The research leading to these results has received funding from the European Research Council under the European Union's Seventh Framework Program (FP7/2013-2018) / ERC Grant Agreement n. 309674.}
}
\begin{document}

\maketitle

\begin{abstract}
Parallel transmission is a very promising candidate technology to mitigate the inevitable radio-frequency (RF) field inhomogeneity in magnetic resonance imaging (MRI) at ultra-high field (UHF). For the first few years, pulse design utilizing this technique was expressed as a least squares problem with crude power regularizations aimed at controlling the specific absorption rate (SAR), hence the patient safety. 
This approach being suboptimal for many applications sensitive mostly to the magnitude of the spin excitation, and not its phase, the magnitude least squares (MLS) problem then was first formulated in 2007. 
Despite its importance and the availability of other powerful numerical optimization methods, the MLS problem yet has been faced almost exclusively by the pulse designer with the so-called variable exchange method. 
In this paper, we investigate various two-stage strategies consisting of different initializations and nonlinear programming approaches, and incorporate directly the strict SAR and hardware constraints. 
Several schemes such as sequential quadratic programming (SQP), interior point (I-P) methods, semidefinite programming (SDP) and magnitude squared least squares (MSLS) relaxations are studied both in the small and large tip angle regimes with RF and static field maps obtained in-vivo on a human brain at 7 Tesla. 
Convergence and robustness of the different approaches are analyzed, and recommendations to tackle this specific problem are finally given. Small tip angle and inversion pulses are returned in a few seconds and in under a minute respectively while respecting the constraints, allowing the use of the proposed approach in routine.
\end{abstract}


\section{Introduction}

One of the main purposes of Ultra High Field (UHF) magnetic resonance imaging (MRI) is to improve spatial resolution, thanks to an increased signal to noise ratio (SNR). The applicability of most MRI sequences however is challenged due to enhanced non-uniformities in the transmit-sensitivity and off-resonance profiles \cite{de2005b1}. 
If not addressed, these can yield zones of important SNR losses across the images, detrimental to diagnosis. 
Over the last few years, a lot of research has been devoted to solve the above-mentioned problems, leading to an assortment of new powerful tools including shaped pulses \cite{moore2012evaluation,boulant2009counteracting}, radio-frequency (RF) shimming \cite{adriany2005transmit} and transmit SENSE parallel transmission (pTX) \cite{katscher2003transmit,zhu2004parallel,grissom2006spatial}. 
While RF shimming may be useful for very specific applications \cite{ellermann2012simultaneous}, due to its versatility pTX so far has proved to be almost indispensable to tackle the RF and static field inhomogeneity problem at UHF for specific absorption rate (SAR) demanding sequences \cite{cloos2012parallel,massire2013design}. 
Due to the large number of degrees of freedom pTX provides, there can be many different SAR patterns for a given RF pulse performance. The goal of the RF pulse designer in general then is to find the pattern that satisfies the SAR guidelines \cite{standardmedical} with the best pulse performance.

Most often, SAR constraints in RF pulse design have either been addressed indirectly through the use of Tikhonov power regularization factors \cite{zelinski2007designing,sbrizzi2011time,deniz2012specific,cloos2010local,sbrizzi2012fast} or by focusing on a more tractable, hence significantly smaller, subset of constraints \cite{graesslin2008minimum,brunner2010optimal}. 
While the first set of approaches requires the tuning of some parameters, the second one generally does not encompass the complexity of the full SAR spatial distribution. 
It was not until recently that the crucial development of the virtual observation points (VOPs) \cite{eichfelder2011local,lee2012local} made the problem considerably more tractable and facilitated the treatment of all the SAR constraints in RF pulse design utilizing pTX. 
In \cite{guerin2013local}, the authors could thereby study more thoroughly the trade-offs between the different constraints and the pulse performance in 2D applications. 
Furthermore, the harder but more beneficial magnitude least squares (MLS) formulation \cite{kerr2007phase,setsompop2008magnitude} of the pulse design problem was to some extent tackled in the latter work either by identifying a suitable target phase and solving the corresponding least squares problem, or by looping over different least squares problems, the so-called variable exchange method \cite{kassakian2006convex}. 
In most cases otherwise, the MLS problem was not addressed. While this can be justified perhaps for spiral k-space trajectories where the rank of the matrix in the linear system is significant, this can lower pulse performance substantially for more sparse trajectories such as spokes \cite{setsompop2008magnitude} and \kT-points \cite{cloos2012kt}. 

The MLS problem in RF pulse design is equivalent to the phase retrieval problem where a complex signal is sought based on the knowledge of magnitude measurements. 
It occurs in other branches of physics such as X-ray crystallography imaging \cite{harrison1993phase}, diffraction imaging \cite{bunk2007diffractive} and microscopy \cite{miao2008extending}. 
The variable exchange method used in the MRI community is identical with the Gerchberg-Saxton algorithm published in 1972 \cite{gerchberg1972practical}. 
Yet, despite the importance of this problem in RF pulse design and its known limitations \cite{waldspurger2012phase}, its use has been ubiquitous in the pulse designers' community \cite{guerin2013local,kerr2007phase,setsompop2008magnitude,cloos2012kt}.  

In this work, we investigate other strategies to solve the MLS problem in 3D, both in the small and the large flip angle (FA) regimes, based on RF field maps measured in-vivo on a human brain at 7 Tesla. The strategies consist of two stages: an initialization and a nonlinear programming approach \cite{Chang2Stage}.
Moreover we incorporate all 10-g and global SAR constraints using VOPs, as well as peak and average power constraints. 
Due to the nonconvexity of the problem, there is no proof that we find the global minimum. 
However, by exploring different techniques such as sequential quadratic programming, interior point methods, semidefinite programming and magnitude squared least squares relaxations, experience can be built and greater confidence can be gained to finally make useful recommendations. 
We first present the general context, by defining the mathematical problem and some computational tricks we used to make the problem more tractable. 
For the sake of clarity and completeness, we then briefly describe the theory behind each technique investigated. 
Their corresponding performance, execution time and robustness are then presented and discussed. 

\section{Theory}

The physical examples we shall work with are non-selective \kT-points pulses \cite{cloos2012kt} with $30^\circ$ and $180^\circ$ target flip angles. The k-space trajectories in both cases were not optimized and consisted respectively of 5 and 7 \kT-points symmetrically located around the centre. The number of \kT-points and the number of channels are respectively called  $N_{k_T}$ and $N_c$ (here equal to 8) throughout. 
The flip angle (MLS) homogenization problem in the small tip angle approximation (STA) \cite{pauly1989k,boulant2012high} is written as:
\begin{equation}
\label{eq:problem}
\min_{x\in \C^p} \| |Ax| - \theta\|_2^2,
\end{equation}
where $\theta$ denotes the target FA (in rads), $A\in \C^{N\times p}$ the spins' dynamics matrix only in the brain region (number of voxels $N = 12 \ 000$, number of columns $p = N_c\times N_{k_T}$), and $x$ the concatenated RF waveforms of the $N_c$ different channels. 
The elements of $A$ are given by
\begin{equation*}
\left.
\begin{array}{ll}
a_{m,(j-1)N_c+n}=&i s B_{1,n}(r_m) \exp(i \langle r_m, k_j\rangle )\\
&\exp\left( i \gamma \Delta B_0(r_m) \left( T-(j-1/2)T_s \right)\right).
\end{array}\right.
\end{equation*}

For computational reasons that will appear shortly, and contrary to general practice \cite{grissom2006spatial,sbrizzi2011time,deniz2012specific,cloos2010local}, the elements for each row are ordered first by channel and then by time. 
Above, $\Delta B_0$ corresponds to the static field offset (in Tesla), $\gamma$ is the gyromagnetic ratio, $B_{1,n}(r_m)$ is the $B_1$ (in $\mu T$) RF field at location $r_m$ for maximum voltage (here 180 Volts) corresponding to the $n^{th}$ channel and $T$ is the total pulse duration. 
The k-space trajectory $k(t)$ is equal to the time-reversed integration of the gradient waveforms to be played during excitation. 
The index $j$ thus labels the \kT-points. Each sub-square pulse is 0.2 ms long for the $30^\circ$ target and 0.5 ms long for the $180^\circ$ target (duration $T_s$ above). 
The scalar $s$ in the expression as a result is the FA (in rads/$\mu T$) achieved for such a sub-pulse with peak RF amplitude of 1 $\mu T$. The time symmetry of the sub-pulses in the low FA regime also implies the $1/2$ correction in the $\Delta B_0$ evolution in the same expression.

To address the SAR constraints, and for the sake of convenience, we use the $Q$ matrices commonly used for SAR calculations \cite{graesslin2008minimum}. 
For a single \kT-point, these are
\begin{equation}
\label{eq:Qmatrix}
Q(r)=\frac{\sigma}{2\rho} \frac{T_s}{TR} \left( E_x^H E_x + E_y^H E_y  + E_z^H E_z\right)(r),
\end{equation}
where $\sigma$ and $\rho$ are the conductivity and density respectively, $TR$ is the repetition time and $E_x$, $E_y$, $E_z$ are the components of the complex electric field row vector at position $r$ along the respective directions for the maximum voltage available on each channel, e.g. $E_x(r) = [E_{x,1}(r), E_{x,2}(r), \hdots, E_{x,N_c}(r)]$. 
The superscript $H$ throughout denotes Hermitian conjugation. 
Here, we take $TR$ such that the overall duty cycle for the $30^\circ$ and $180^\circ$ pulses are 10 \% and 0.25 \% respectively. For a single \kT-point the SAR at location $r$ for a pulse shape $x$ that way is $\text{SAR}(r)=x^H Q(r)x$. 
Equation \eqref{eq:Qmatrix} then is simply averaged over 10-g of contiguous tissue to obtain the corresponding $Q_{10g}(r)$ matrices. 
Strictly speaking, thus the number of 10-g SAR constraints is equal to the number of voxels in the head model. In the human head model we describe later on, it is roughly equal to $37 \ 000$. 
To make the problem more tractable, we have used the compression model described in \cite{eichfelder2011local}. This model returns a set of VOPs with corresponding $Q_{VOPs}$ matrices which guarantees that $\exists i\in VOPs$ so that 
\begin{equation}
\label{SARineq}
SAR_{10g,i}\leq \max_{r\in \Omega}(SAR_{10g}(r))\leq SAR_{10g,i}+\epsilon_G SAR_{global}, 
\end{equation}
where $\Omega$ denotes the ensemble of all voxels.
Respecting the constraints over the VOPs with a certain pre-defined tolerance hence guarantees to satisfy the constraints in every voxel. The number of VOPs, here denoted as $N_{VOPs}$ throughout for the sake of generality, depends on $\epsilon_G$, the model and the experimental set-up. 
In this study, we have taken $\epsilon_G=1$, which returned a set of 490 VOPs. 
A matrix $Q_G$ was also used to deal with the constraint on global SAR. 
In addition we incorporate peak power and average power constraints for each channel (again taking into account the duty cycles). Finally the optimization problem for the $30^\circ$ target FA can be summarized as the following:
\begin{equation}
\label{eq3}
\mmin{f(x):=\| |Ax| - \theta \|_2^2}{
x\in \C^p, \\
c_i(x) \leq 10 \ \mathrm{W/kg}, \ i \in \{1,\hdots, N_{VOPs}\}, \\
c_G(x) \leq 3.2 \ \mathrm{W/kg}, \\
c_{pw,k}(x) \leq 10 \ \mathrm{W}, k\in \{1, \hdots, N_c\}, \\
c_{A,j}(x) = |x_j|^2 \leq 1, j\in \{1, \hdots, N_c \times N_{k_T}\}.
}
\end{equation}
The functions $c_i$, $c_G$, $c_{A,j}$ and $c_{pw,k}$ are quadratic functions. They denote the 10-g SAR constraints over the VOPs (calculated with $Q_{VOPs}$), the global SAR constraint (calculated with $Q_{G}$), the amplitude and the average power for the $k^{th}$ channel (here taken as 10 $\mathrm{W}$) constraints  respectively. 
The values for the SAR constraints above correspond to the guidelines issued by the International Electrotechnical Commission (IEC)  \cite{standardmedical}. 
For the inversion pulse, the FA in the objective function in  \eqref{eq:problem} is no longer expressed as a linear relationship but is obtained via the nonlinear function $bl$ (for Bloch):
\begin{equation}
\label{eq4}
\mmin{f(x):=\| |bl(x)| - \theta \|_2^2}{
x\in \C^p, \\
c_i(x) \leq 3 \ \mathrm{W/kg}, \ i \in \{1,\hdots, N_{VOPs}\}, \\
c_G(x) \leq 1 \ \mathrm{W/kg}, \\
c_{pw,k}(x) \leq 2 \ \mathrm{W}, k\in \{1, \hdots, N_c\}, \\
c_{A,j}(x) \leq 1, j\in \{1, \hdots, N_c\times N_{k_T}\}.
}
\end{equation}

Note that the SAR and average power thresholds here are different than in problem \eqref{eq3}. 
This is because inversion pulses are often used in combination with many additional low FA pulses and that both participate in yielding an effective SAR pattern and an average power. 
An important example is the MPRAGE sequence, where the duty cycle of that pulse indeed is on the order of 0.25 \% ($TR$ $\simeq$ 2s). 
This is why the SAR thresholds above have been set, arbitrarily, about 3 times less than the ones recommended by the IEC, keeping in mind that this pulse would likely be used in combination with other small FA pulses.

Both problems \eqref{eq3} and \eqref{eq4} are nonconvex optimization problems with nonlinear constraints. 
All techniques we have explored to solve them make use of the values of the objective function, the constraints and their respective gradients. We detail briefly how these values were calculated later on. 
All CPU times provided in this paper are for a workstation equipped with an Intel Xeon E5-2620 processor, 16 GB of RAM and using either the Matlab R2013a (The Mathworks, Natick, MA, USA) or the Knitro (Ziena optimization LLC, Evanston, IL, USA) software depending on the algorithms. 
The Bloch integrations needed for problem \eqref{eq4} were performed with CUDA on a Graphics Processing Unit (GPU) NVIDIA (Santa Clara, CA, USA)  Tesla C1060 card.

\section{Methods}

\subsection{Head model, $B1$ and $\Delta B0$ maps.}

\begin{figure}[htb]
 \centering
 \centerline{
  \subfigure[]{\label{fig:headModela}\includegraphics[width=4cm]{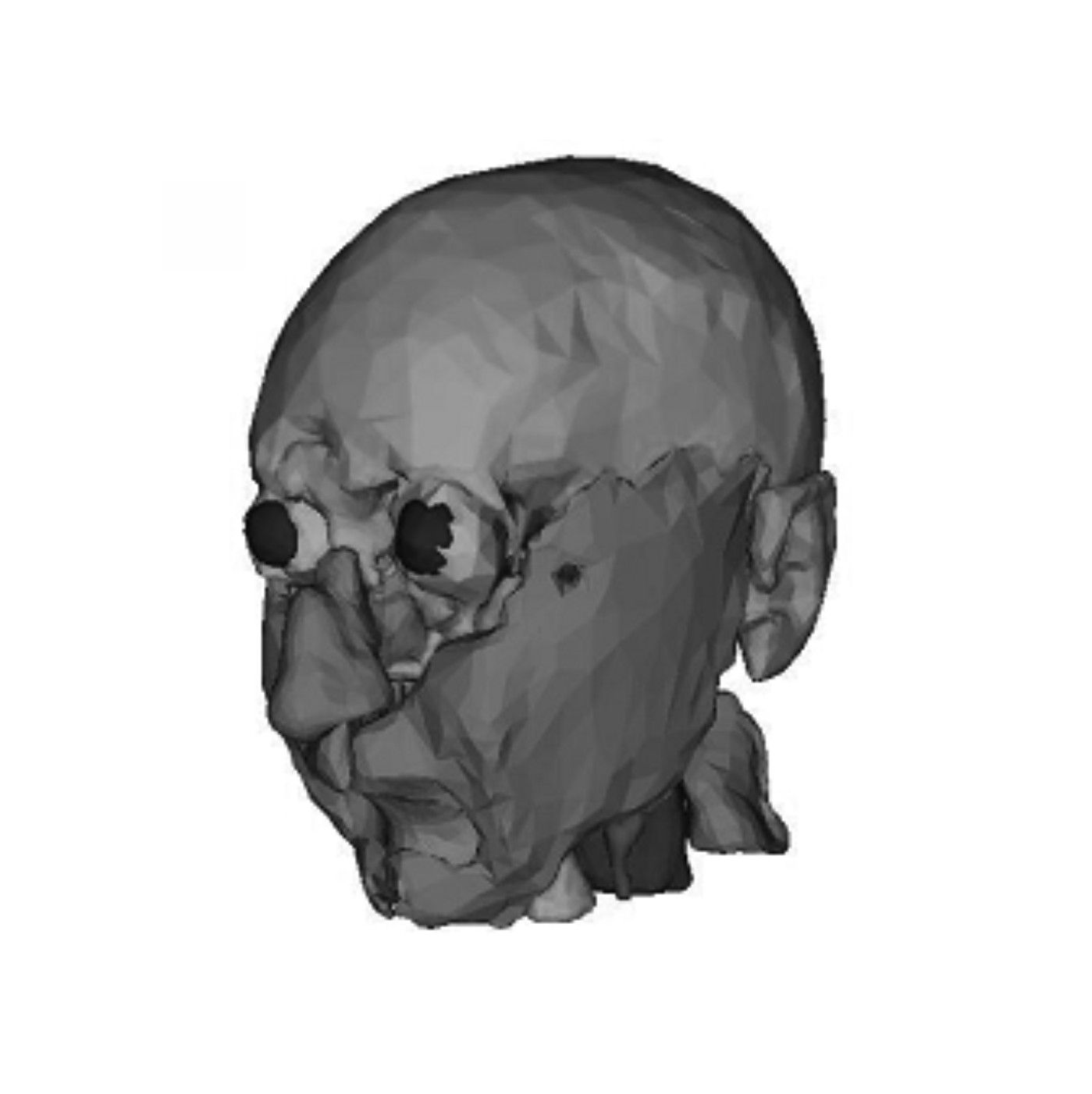}}
  \subfigure[]{\label{fig:coil}\includegraphics[width=4cm]{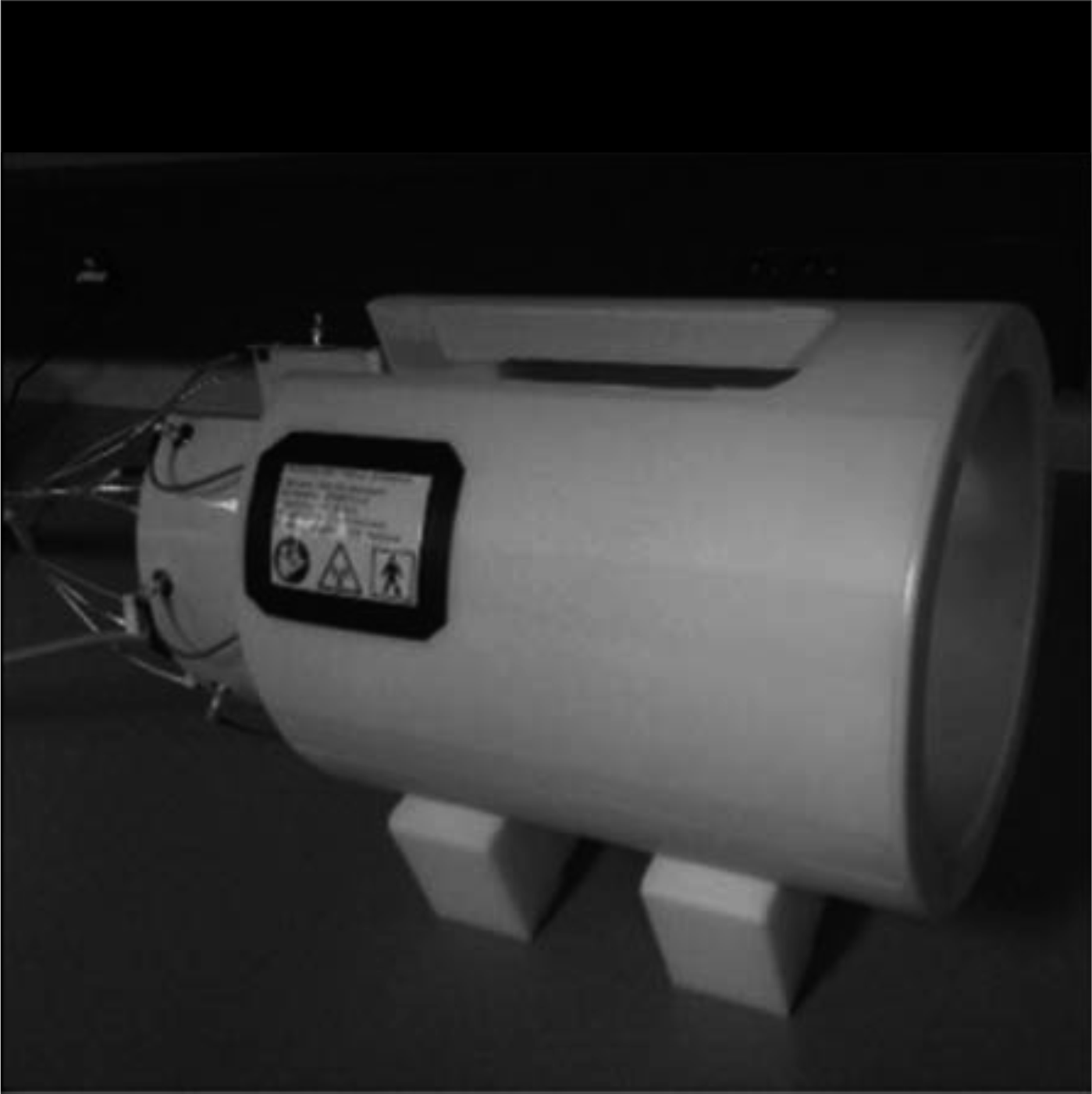}}
  }
\caption{Head model \subref{fig:headModela} and 8-channel pTX coil \subref{fig:coil}. Only subcutaneous tissues are shown for
illustration in a). Both models of the coil and the head were imported into HFSS to perform finite element electromagnetic simulations. Electric fields were returned for each voxel of the head for a given input power on each transmit element. The electric fields were then used to build the $Q$ matrices in \eqref{eq:Qmatrix} and the VOPs \cite{eichfelder2011local,lee2012local}}.
\label{fig:HeadModel}
\end{figure}

The numerical head model we used for the SAR calculations is the one described in \cite{massire2012thermal} whose surface-based representation was obtained from the voxel-based model reported in \cite{makris2008mri}. Such a representation was a necessary step to be able to run finite element electromagnetic simulations in HFSS (Ansys, Canonsburg, PA, USA) with our 8-channel pTX coil tuned and matched at 297 MHz to correspond to the proton Larmor frequency at 7 Tesla. Electric fields computed by HFSS were projected onto a $5\times 5\times5 \ \mathrm{mm}^3$ Cartesian grid and used to build up the $Q$ matrices of \eqref{eq:Qmatrix}. The head model (illustrated in Fig. \ref{fig:headModela}) contains 20 anatomical structure entities with corresponding electric properties, which added up to around $37 \ 000$ voxels. The RF and static field maps were acquired at 7 Tesla on a healthy adult volunteer, for which approval was obtained from our institutional review board. The same dataset was used in some of our earlier work \cite{cloos2012kt}.
 For both RF transmission and reception, a home-made transceiver-array head coil was used (Fig. \ref{fig:coil}), which consists of eight stripline dipoles distributed every $42.5^\circ$ on a cylindrical surface of 27.6-cm diameter, leaving a small open space in front of the subject's eyes. Relative $B_1$ maps were first acquired using 8 small tip angle FLASH sequences \cite{van2007calibration}, with a small $T_1$ nonlinearity correction. Two reference actual FA acquisitions \cite{yarnykh2007actual} were then acquired to convert the previous data into absolute maps, with additional echoes in the first repetition time to monitor the $\Delta B_0$ evolution. Matrix size was $48\times 48 \times 32$ with isotropic resolution of 5 mm. These data allowed the encoding of the $A$ matrix in \eqref{eq:problem}. Due to the relative smoothness of the $B_1$ and $\Delta B_0$ maps, significant differences in homogeneity when dealing with higher resolution maps are not expected. Finally, although the direct link between the resolution and the number of VOPs is not known to the authors, a higher resolution can lead to increased accuracy of the SAR results but at the possible expense of computational speed.  

\subsection{Algorithms}

The nonconvex problems defined by \eqref{eq3} and \eqref{eq4} are solved using a two stage optimization strategy, where in the first stage an initial vector $x$ is determined and used as an input to the second stage, a nonlinear programming algorithm. The techniques we investigated for the former stage are random initialization, the Gerchberg-Saxton algorithm \cite{gerchberg1972practical} and Semidefinite Relaxation (SDR) \cite{waldspurger2012phase}. For the nonlinear programming stage, we studied the algorithms proposed by the optimization toolbox of Matlab, i.e. the active-set (A-S), Sequential Quadratic Programming (SQP), interior-point (I-P) algorithm, as well as some variants proposed by the Knitro software allowing a user-supplied Hessian of the Lagrangian.

\subsubsection{Initialization phase}

\paragraph{Random initialization}

To increase our chances to find the global minimum, 500 random vectors (uniform distribution) were generated and scaled to satisfy initially all constraints. These inputs were then fed to each different nonlinear programming algorithm. In the large FA case however, and when the Hessian of the Lagrangian was user-supplied, only 50 random vectors were tried due to the long execution time.  

\paragraph{Gerchberg-Saxton}

Also called variable exchange method \cite{kassakian2006convex}, this algorithm is inspired from the alternating projection strategy \cite{bauschke2011convex} where the problem \eqref{eq:problem} is reformulated in a way that does not involve the absolute value function by explicitly splitting amplitude and phase contributions, $\mathop{diag}(\theta)$ and $u$, with $u$ satisfying the unity modulus constraint $|u_i|=1$ for $i\in \{1,\hdots,N\}$. The problem is thus written as
\begin{equation}
\label{eq6}
\mmin{\|Ax-\mathop{diag}(\theta) u\|_2^2}{u\in \C^N, \\ |u_i|=1,\\ x\in \C^p, \\ }
\end{equation}
where the optimization procedure is iterated over both variables $u\in \C^N$ and $x\in \C^p$. For fixed $u$, $x$ can be found by solving the unconstrained linear least squares problem:
\begin{equation}
\label{eq7}
x=A^\dag \mathop{diag}(\theta) u,
\end{equation}
where $A^\dag$ is the Moore-Penrose pseudo-inverse of $A$. Despite the crude approximation thereby made for the large FA, where the relation between the pulse shape and the FA is nonlinear, it is an easily implementable and quick technique able to return a nonrandom and reasonable guess. The Gerchberg-Saxton (G-S) algorithm projects the current $\mathop{diag}(\theta)u^h$ on the image of $A$ using the orthogonal projector  $AA^\dag$ and correspondingly updates its phase for the new iterate (which is just equivalent to adjust the new phase to the one of $Ax^h$). This procedure is repeated until some convergence criteria are satisfied, e.g. when the cost function varies less than by 1 \% from one iteration to the next. Nevertheless, and as the problem is nonconvex in $u$, this alternating projection method is known to stall in local minima when there are no constraints \cite{waldspurger2012phase}. This method also requires an initial phase distribution in the vector $u$, which we took to be the one of the circularly polarized mode 
obtained by aligning the phases of the RF field maps at the center of the brain. Finally, the G-S algorithm the way we have presented it so far generates only one candidate $x$. To generate different candidates, we solved instead the following problem
\begin{equation}
\label{eq8}
\mmin{\|Ax-\mathop{diag}(\theta) u\|_2^2 + \lambda \|x\|_2^2}{u\in \C^N, \\ |u_i|=1,\\ x\in \C^p, \\ },
\end{equation}

where $\lambda$ is a regularization parameter which was varied logarithmically from 1 to 300 and $10 \ 000$ for the small and large FA targets respectively (500 steps). The range in the latter case indeed was increased as it appeared it generated more variability in the returned input vectors. The G-S algorithm in this case is implemented in the same way this time by using $x=\tilde{A}_\lambda \mathop{diag}(\theta) u$ where $\tilde{A}_\lambda=(A^H A+\lambda \Id)^{-1}A^H$ \cite{setsompop2008magnitude}. Cross-correlations between the corresponding results were calculated to assess their similarities.

\paragraph{Semidefinite Relaxation}

The problem \eqref{eq8} can be expressed as a quadratic programming problem (QP) with variable $u$ by inserting  $x=\tilde{A}_\lambda \diag(\theta)u$ in the same equation and setting the positive definite Hermitian matrix  $M=\diag(\theta)(\Id-A\tilde{A}_\lambda)\diag(\theta)$ \cite{waldspurger2012phase},
\begin{equation}
\label{eq9}
QP(M):= \mmin{u^HMu}{u\in \C^N, \\ |u_i|=1, i\in \{1,\hdots, N\}.}
\end{equation}

This problem is equivalent to a Semidefinite Program (SDP) in $U=uu^H \in \mathcal{H}_N$: 
\begin{equation}
\label{eq10}
SDP(M):= \mmin{\mathop{Tr}(UM)}{U\in \mathcal{H}_N, \\ \diag(U)=1, \\ U\succeq 0, \mathop{Rank}(U)=1.}
\end{equation}
The Semidefinite Relaxation (SDR) is obtained by dropping the nonconvex rank constraint and is known to have theoretical guarantees about the global minimum when there are no additional constraints. When the solution has rank one, the relaxation is tight and the vector $u$ is an optimal solution to the problem defined by \eqref{eq9} \cite{waldspurger2012phase}. If the solution has rank larger than one, a normalized leading eigenvector of $U$ is used as a candidate solution. In practice, the semidefinite programing solvers are rarely designed to handle complex matrices. Therefore the complex programs are often reformulated using the linear transformation $T$ that maps Hermitian complex matrices $\mathcal{H}_N$ in  to real semi-definite positive matrices in $S_{2N}$ \cite{goemans2001approximation},
\begin{equation}
\label{eq11}
T(M) = \left[\begin{array}{cc} Re(M) & -Im(M) \\ Im(M) & Re(M) \end{array}\right].
\end{equation}
We used the code available at (\url{http://www.cmap.polytechnique.fr/scattering/code/phaserecovery.zip}), which is a block-coordinate descent method, to solve this new problem. Because solving this problem using this approach took several hours given the size of the matrices, we tried only three different values of $\lambda$ to generate different initial guesses (1, 100, and 300 for the small FA and 1, 100 and $10 \ 000$ for the large FA). These candidates likewise were then fed as initial starting points to the nonlinear programming algorithms we investigated.

\subsubsection{Nonlinear programming}

Some of the most successful large scale algorithms for generally constrained nonlinear optimization fall into one of two categories \cite{nocedal1999wright}:  active-set sequential quadratic programming methods and interior-point or barrier methods. Active-set methods can quickly generate a good working set of active constraints, i.e. the ones that satisfy the constraint equalities in Problems (4) and (5), and then perform a minimization on the smaller dimensional subspace generated by this set of linearized and active constraints which are iteratively updated. The constraints that are estimated to be nonactive at the solution point are simply disregarded. Interior-point methods on the other hand always attempt enforcing all the constraints by using barrier penalty functions which progressively vanish as the number of iterations increases. The efficiency and the scaling of these schemes with respect to the size of the problem  heavily depend on the particular problem of interest \cite{hei2008numerical}. We now briefly detail these methods.

\paragraph{Active-set methods}

They consist of a (Quasi-)Newton procedure where at each step the problem is approximated by a QP with the constraints linearized \cite{hei2008numerical}. Let $QP(x^h,H_h)$ denote the following problem:
\begin{flushleft}
\begin{equation}
\label{eq12}
\mmin{\frac{1}{2} d^H H_h d + \nabla f(x^h)^H d}{d\in \C^p, \\
c_i(x^h) + \nabla c_i(x^h)^Hd \leq 10 \ \mathrm{W/Kg}, i\in \{1,\hdots, N_{VOPs}\}, \\
c_G(x^h)+\nabla c_G(x^h)^Hd \leq 3. 2 \ \mathrm{W/Kg}, \\
c_{pw,k}(x^h) + \nabla c_{pw,k}(x^h)^Hd \leq 10 \ \mathrm{W}, k\in \{1,\hdots,N_c\}, \\
c_{A,j}(x^h) + \nabla c_{A,j}(x^h)^H d\leq 1 , j\in \{1, \hdots, N_c\times N_{k_T}\},}
\end{equation}
\end{flushleft}
where $d=x-x^h$ is the unknown step vector to take at the $h^{th}$ iteration and $H_h$ is the Hessian of the Lagrangian updated in the Matlab implementation via the Broyden-Fletcher-Goldfarb-Shanno (BFGS) method \cite{hei2008numerical}. The basic principle of the A-S algorithm is to solve problem \eqref{eq12} and estimate the active constraints via a calculation of the Lagrange multipliers. At each iteration the constraints believed to be nonactive at the solution point are simply disregarded (but can be reincorporated at a later iteration). For those that are active, the linear constraints above allow to find a subspace tangent to these in which a search direction is determined. The SQP implementation of Matlab is similar to A-S \cite{han1977globally}, the main differences being: the strict feasibility with respect to bounds, this is beneficial when the objective function or the nonlinear constraints functions are undefined or are complex outside the region constrained by the bounds; feasibility routines are reformulated; SQP algorithm is more robust to non-double results, hence it tries to ensure numerical convergence; the linear algebra routines are refactored in order to be more efficient in memory usage and speed.

\paragraph{Interior-Point Methods}

The I-P approach consists of approximating the problem by a sequence of equality constrained problems that are easier to solve than the original inequality-constrained problem \cite{nocedal1999wright}. It reads:
\begin{equation}
\label{eq13}
\mmin{ f(x)  - \mu_k \sum_{l\in \mathcal{J}} \log(s_l) }{
c_i(x) + s_i = 10 \ \mathrm{W/Kg}, i\in \{1,\hdots, N_{VOPs}\}, \\
c_G(x)+ s_G = 3.2 \ \mathrm{W/Kg}, \\
c_{pw,k}(x) + s_{pw,k} = 10 \ \mathrm{W}, k\in \{1,\hdots,N_c\}, \\
c_{A,j}(x) + s_{A,k} d= 1 , j\in \{1, \hdots, N_c \times N_{k_T}\}, \\
s_l\geq 0, l\in \mathcal{J},
}
\end{equation}
where $\mathcal{J}$ is the set of constraints, $s$ is a vector of slack variables and $\mu_k>0$ is the sequence of the barrier parameter with $\lim_{k\rightarrow +\infty} \mu_k=0$. There are two main classes of I-P methods \cite{nocedal1999wright}. The first class can be viewed as direct extensions of I-P methods for linear and quadratic programming. They use line searches to enforce convergence, computing the descent steps by solving a system of equations corresponding to the Karush-Kuhn-Tucker \cite{nocedal1999wright} conditions; they are also called I-P with direct step. The methods in the second class use a quadratic model to define the step and incorporate a trust-region constraint to provide stability; they are also called I-P with conjugate gradient (CG) step \cite{nocedal1999wright}. Matlab exploits both classes in its implementation. In both I-P and A-S methods, the number of iterations, i.e. the number of times problem \eqref{eq12} or problem \eqref{eq13} was solved, was set equal to 60 which was enough in this particular problem to reach convergence towards a local minimum.

\subsection{Implementation}

\subsubsection{Calculation of the objective function, the SAR constraints and their respective gradients.}

Given a candidate solution $x$ the objective function in the small tip angle regime is calculated easily by using \eqref{eq3}, where most of the computation consists of multiplying the matrix $A$ with the vector $x$ and taking an $l^2$-norm. Because most optimization algorithms do not deal directly with complex variables, we first split the vector $x$ into two parts: $\ds \begin{bmatrix} Re(x) \\ Im(x) \end{bmatrix}$. The gradient $\ds \nabla f\begin{pmatrix} Re(x) \\ Im(x) \end{pmatrix}$could be obtained by blind finite difference with $\ds \frac{\partial f}{\partial x_j}(x) \simeq \frac{f(x+\epsilon e_j) -f(x)}{\epsilon}$, where $e_j$ is an elementary vector (1 at one position and zero elsewhere) but this would imply calculating the same kind of matrix-vector product for each degree of freedom, i.e. $2 \times N_c \times N_{kT}$ times (here 80 for the small FA pulse). By denoting $a_j$ the $j^{th}$ column of the matrix $A$, instead by linearity one has $A(x+\epsilon e_j)= Ax + \epsilon a_j$, for the first half of the gradient vector and 
$Ax+i\epsilon a_{j-p}$, for the second half, where $i=\sqrt{-1}$. The product $Ax$ thereby is performed only once. This saves many unnecessary calculations. At the end of the procedure, the normalized root mean square error (NRMSE) ${\sqrt{\frac{1}{N} \sum\limits_{i=1}^{N} (\theta_i - \theta)^2}/\theta}$ is provided. In the large tip angle regime, the objective function is calculated by using \eqref{eq4}, where most of the computation consists of carrying a full Bloch simulation over all voxels and taking an $l^2$-norm. Its gradient in this case is evaluated via finite differences. The evaluation of the objective function, thus a crucial step to make this approach feasible in routine, was performed by using CUDA and a GPU card.  

For one \kT-point, the SAR for one VOP for a pulse shape $x$ is equal to $c_{VOP}(x)=x^H Q_{VOP}x$. 
For $N_{k_T}$ \kT-points back to back, the result conveniently becomes $c_{VOP}(x)=x^H (\Id_{N_{k_T}} \otimes Q_{VOP}) x$ thanks to the ordering of the elements in the $A$ matrix, where $\Id_{N_{k_T}}$ is the Identity matrix of size $N_{k_T}$ and $\otimes$ is the Kronecker product. 
If $N_{k_T}$ is not small, the matrix sandwiched between the $x$ vectors is sparse and can be declared as such to speed up the computations. 
Most importantly, the gradient of the SAR hence has the following analytical expression: 
\begin{equation*}	
\nabla c_{VOP}\begin{pmatrix} Re(x) \\ Im(x) \end{pmatrix} = 
2 \begin{bmatrix} 
Re((\Id_{N_{k_T}}\otimes Q_{VOP}) x) \\
Im((\Id_{N_{k_T}}\otimes Q_{VOP}) x) 
\end{bmatrix}.
\end{equation*}
Although the VOPs considerably reduce the size of the problem, doing a loop over all of them to calculate the SAR would be suboptimal, especially for less compressed models. If $I_c$ is the matrix whose columns are the $N_T$-time point waveforms of the different channels, for 100 \% of duty cycle the SAR at one spatial location can be calculated using the following formula \cite{graesslin2012specific}: 
\begin{align}
SAR&=\sum_{m=1}^{N_c} \sum_{n=1}^{N_c} Q_{m,n} \overline{I_{c,m}^H I_{c,n}},  \\
&= \mathbb{1}^T\left( Q \odot \left( \frac{1}{N_T} I_c^H I_c\right)\right) \mathbb{1},
\end{align}

where $\mathbb{1}^T = [1, \hdots, 1] \in \R^{N_c}$, the operator on the right hand side denotes the Hadamard product and the bar denotes a time average. For one voxel, this formula already saves looping over the different time points. To calculate the SAR value over all the VOPs, we simply compute:
\begin{equation}
\label{eq14}
SAR
= \mathbb{1}^T \left( Q_{cat} \odot  \left[ \underbrace{\overline{I_c^H I_c}, \hdots,  \overline{I_c^H I_c}}_{N_{VOPs} \ \textrm{times}} \right] \right) \begin{bmatrix} \mathbb{1} & \hdots & 0\\ \vdots & \hdots & \vdots & \\ 0 & \hdots & \mathbb{1} \end{bmatrix},
\end{equation}

where  $Q_{cat}=[Q_1, \hdots, Q_{N_{VOPs}}]$.
The matrix on the far right contains $N_{VOPs} \times N_c$ rows and should be declared as sparse (1.6 \% of the elements are nonzeros in our case). Each column of that matrix contains 8 consecutive ones which isolate a $Q$ matrix. This way, all SAR values are computed in one shot with optimized algebra routines and with no \textit{for loops}, often time-consuming in Matlab implementations. All matrices besides $I_c^H I_c$ are built once and for all throughout. Finally, because analytical formulas likewise exist for the gradients of the peak and average power constraints, they can be calculated very efficiently as well.

\subsubsection{MSLS problem and  the Hessian of the Lagrangian.}

Since the objective function defined by \eqref{eq:problem} is not everywhere differentiable, gradient-based methods are not well defined and can cause problems. For that reason we attempted to solve the original MLS problem by using a variant that is different but closely related, i.e. the Magnitude Squared Least Squares (MSLS) problem \cite{kassakian2006convex}, still under the same constraints as in the MLS problems (\ref{eq3}) and (\ref{eq4}):
\begin{equation}
\label{eq15}
\min_{x\in \C^p} f(x):= \| |g(x)|^2 -\theta^2 \|_2^2,
\end{equation}
where $g(x) = Ax$ and $g(x)=bl(x)$ for the small and large tip angle regimes respectively. In the small FA regime the gradient and the Hessian of the Lagrangian are determined analytically \cite{kassakian2006convex} and supplied at moderate cost to the Knitro solver (Matlab does not accept a user-supplied Hessian for the A-S and I-P methods). Like for SDP, in order to work with real variables, the linear transformation defined by \eqref{eq11} is applied to $A$, $M=T(A)$ and setting $M_j=[col_j(M), col_{p+j}(M)]$, the gradient and the Hessian of the objective function are \cite{kassakian2006convex}
\begin{equation*}
\nabla f(x) = 4\sum_{i=1}^N (x^H M_iM_i^Hx - \theta^2)M_iM_i^Hx 
\end{equation*}
and 
\begin{align*}
\nabla^2 f(x) =& 4 \sum_{i=1}^N 2M_iM_i^H xx^H M_iM_i^H \\ &+(x^H M_iM_i^H-\theta^2) M_iM_i^H 
\end{align*}
respectively. 
For the large tip angle regime the Hessian of the function, $\ds \nabla^2 f\begin{pmatrix} Re(x) \\ Im(x)\end{pmatrix}$, was obtained by blind finite differences.
Second derivatives for the SAR and power constraints were trivially added to this contribution to return the Hessian of the Lagrangian. Based on this reformulation we investigated likewise the A-S and I-P methods. Knitro on the other hand provides the user with the choice of using either the direct or CG approach for the latter, so that both were tried. Despite the use of a GPU to calculate the objective function and the finite differences in the large tip angle regime, calculating explicitly the Hessian significantly increased the numerical burden compared to the much faster, but approximate, BFGS update method. However the goal in using the MSLS problem reformulation was to investigate potential gains in the returned RF pulse performance (NRMSE) by exploiting full knowledge of the Hessian matrix of the Lagrangian. Because of the longer duration of this implementation (around 20 minutes per run), only 50 random and 50 G-S initializations were performed in the large FA case (as opposed to 500 for the small FA). Semidefinite relaxation, as presented above, was also used to supply three different initial guesses to the Knitro solver. Knowledge of the Hessian allowed faster convergence in terms of number of iterations, which was therefore set to 30. 

\subsubsection{Summary of the different approaches.}

We summarize the different schemes investigated in this paper for the reader's convenience. Initializations were performed using: random draws, the G-S algorithm and SDR. The problem of interest is the MLS problem defined in \eqref{eq3} and \eqref{eq4}. Both A-S and I-P methods were attempted using a Hessian of the Lagrangian updated by the BFGS method for Matlab and a user-supplied one for Knitro (determined analytically for the small FA problem and via finite differences for the large FA one). The MSLS variant was used in the latter case.   

\section{Results}

\begin{figure}[htb]
 \centering
 \centerline{
  \subfigure[Low flip angle regime]{\label{fig:STAbox}\includegraphics[width=4.5cm]{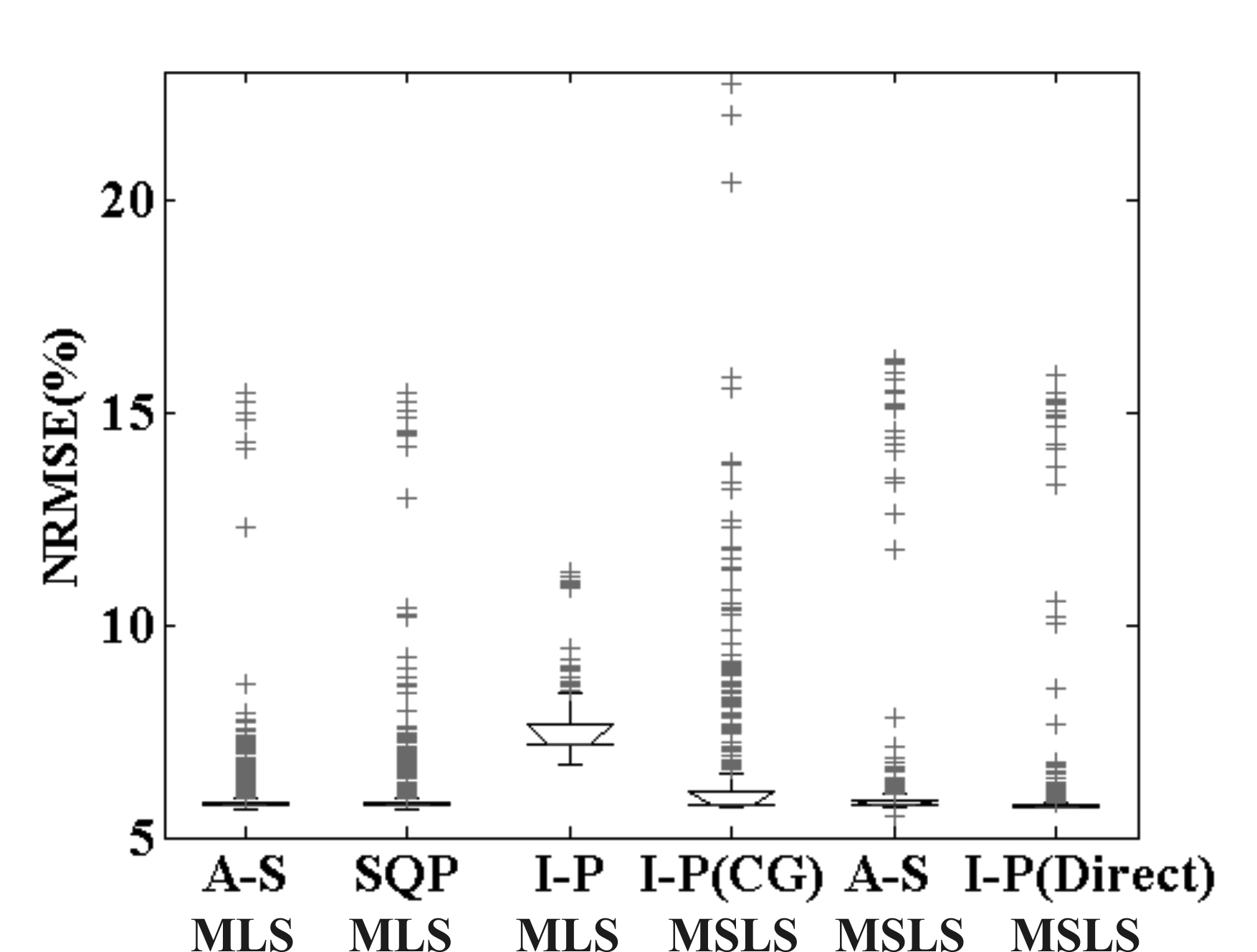}}
  \subfigure[High flip angle regime]{\label{fig:LTAbox}\includegraphics[width=4.5cm]{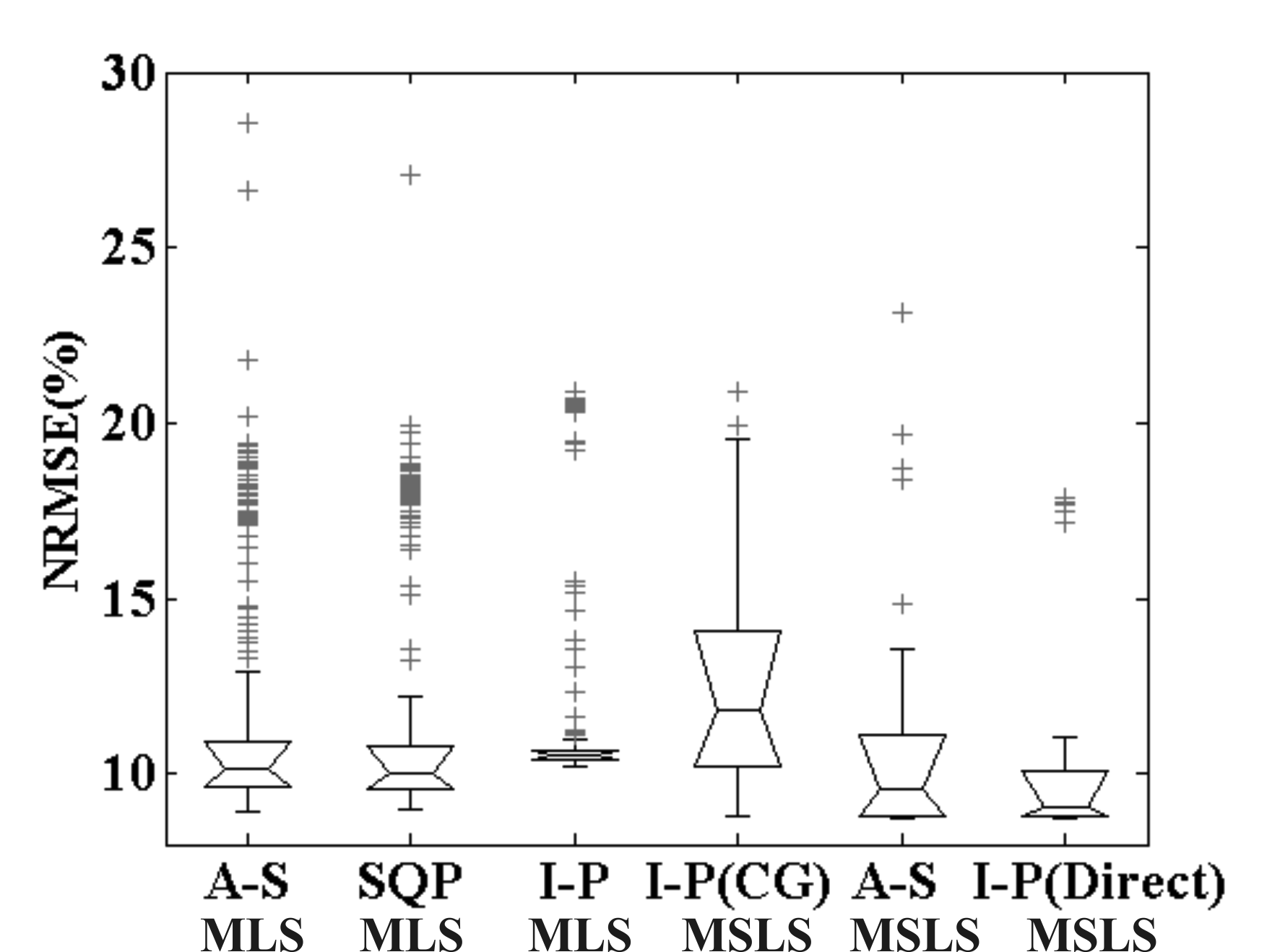}}
  }
\caption{ Box plots obtained for the small \subref{fig:STAbox} and the large tip angle regimes \subref{fig:LTAbox}. The small horizontal lines correspond to the lowest and highest data within 1.5 interquartile range below the first and above the third quartiles respectively. The crosses are the outliers. The plots were generated with 500 random samples for all approaches except for the MSLS variant in the large FA regime (50 random samples).}
\label{fig:boxplot}
\end{figure}

The cross-correlation among the different input vectors varied between 0.5 and 1 for the random draws, while the one among the vectors generated by the G-S algorithm varied between 0.65 and 1, which indicates a significant variability in the input states. Fig.\ref{fig:boxplot} provides the box plots showing the variability of the final NRMSEs with the random initialization procedure combined with the different algorithms. In the small FA regime one can observe a smaller variability within the 1.5 interquartile range for the A-S methods compared to the I-P ones, except for the I-P direct approach in the MSLS formulation. The result however is opposite for the high FA problem with a higher robustness achieved for the I-P method in the original MLS formulation.

Table \ref{tab1} summarizes the best NRMSEs obtained for the small and large FAs and for each algorithm, along with their execution times (not including the initialization routine which, besides the SDR method, takes negligible time). The constraints before and after each algorithm are also reported in the second table, the initializations being the ones leading to their best respective results. The SDR initialization method returned matrices whose ranks were always far larger than 1 so that no statements about global optimality (without constraints) could be made. 
The differences between the A-S and SQP implementations of Matlab were minor. The SQP program was slightly slower but guaranteed at the end the strict respect of all the constraints. The A-S technique could on the other hand return a result where the constraints could be, very slightly, violated. Execution time naturally varied significantly depending on the evaluation method of the Hessian (analytical, BFGS or finite differences), the evaluation of the objective function and its gradient (STA, Bloch simulation, finite differences), the algorithm and its starting point. Solving the MSLS problem with Knitro in the large FA regime for instance takes nearly 20 minutes due to the evaluation of the Hessian via finite differences, even when using a GPU card. Solving the MLS problem using the A-S approach and the BFGS update method on the other hand takes 6 and 47 seconds for the small and large FAs respectively.

\begin{table*}
\caption{Best NRMSE in \% and execution time in seconds (in parenthesis) obtained for each initialization/algorithm combination for the small FA (SFA) and large FA (LFA) regimes.}
\label{tab1}
\centering
\begin{tabular}{cc|c|c|c|c|c|c|c|}
  \cline{2-8}
  & \multicolumn{1}{ |c }{\backslashbox { Initialization }{ Algorithm }}        & \multicolumn{3}{|c|}{MLS (Matlab)}        & \multicolumn{3}{c|}{MSLS (Knitro)}   \\
  \cline{2-8}\\[-0.8em]
  \cline{3-8}
   & & A-S & SQP & I-P direct \& CG & A-S  & I-P direct & I-P CG \\
  \hline
  \multicolumn{1}{ |c| }{\multirow{3}{*}{\rotatebox[origin=c]{90}{SFA}}} 
   & Random  &  5.67 (6)         & 5.67 (7)  &  6.70 (9)   &  5.68 (40)       & 5.68 (35)  & 5.73 (34)\\
   \multicolumn{1}{ |c| }{} & G-S     & 5.70 (6)          & 5.71 (8) &  7.24 (8)&  5.68 (47)      & 5.73 (33)    & 5.69 (33)\\ 
   \multicolumn{1}{ |c| }{} & SDR     &  \textbf{{5.65}} (8) & 5.65 (9) &  12.67 (11) &  5.77 (40)   & 6.42 (37)  & 10.72 (36)\\ 
  \hline
  \multicolumn{1}{ |c| }{\multirow{3}{*}{\rotatebox[origin=c]{90}{LFA}}}
   & Random  & 8.94 (49) & 9.00 (78) & 10.22 (44)  &  8.83 ($1\ 369$)  & 8.76 ($1\ 124$)         & 9.07 ($1 \ 126$)     \\ 
   \multicolumn{1}{ |c| }{} & G-S     & 8.90 (47)  & 8.91 (77)& 10.20 (38)  &  \textbf{{8.72}} ($1 \ 351$)  & 8.74 ($1\ 119$)         & 8.83 ($1\ 111$)     \\ 
   \multicolumn{1}{ |c| }{} & SDR     & 9.42 (49)  & 9.49 (48)& 10.34 (44)  &  9.57 ($1 \ 357$)  & 9.03 ($1\ 116$)         & 12.54 ($1\ 117$)     \\ 
  \hline
\end{tabular}
\end{table*}

\begin{figure}[htb]
\renewcommand{\arraystretch}{1.3}
 \centering
 \centerline{
  \subfigure[Low flip angle regime]{\label{fig:costFuncSTA}\includegraphics[width=4.5cm]{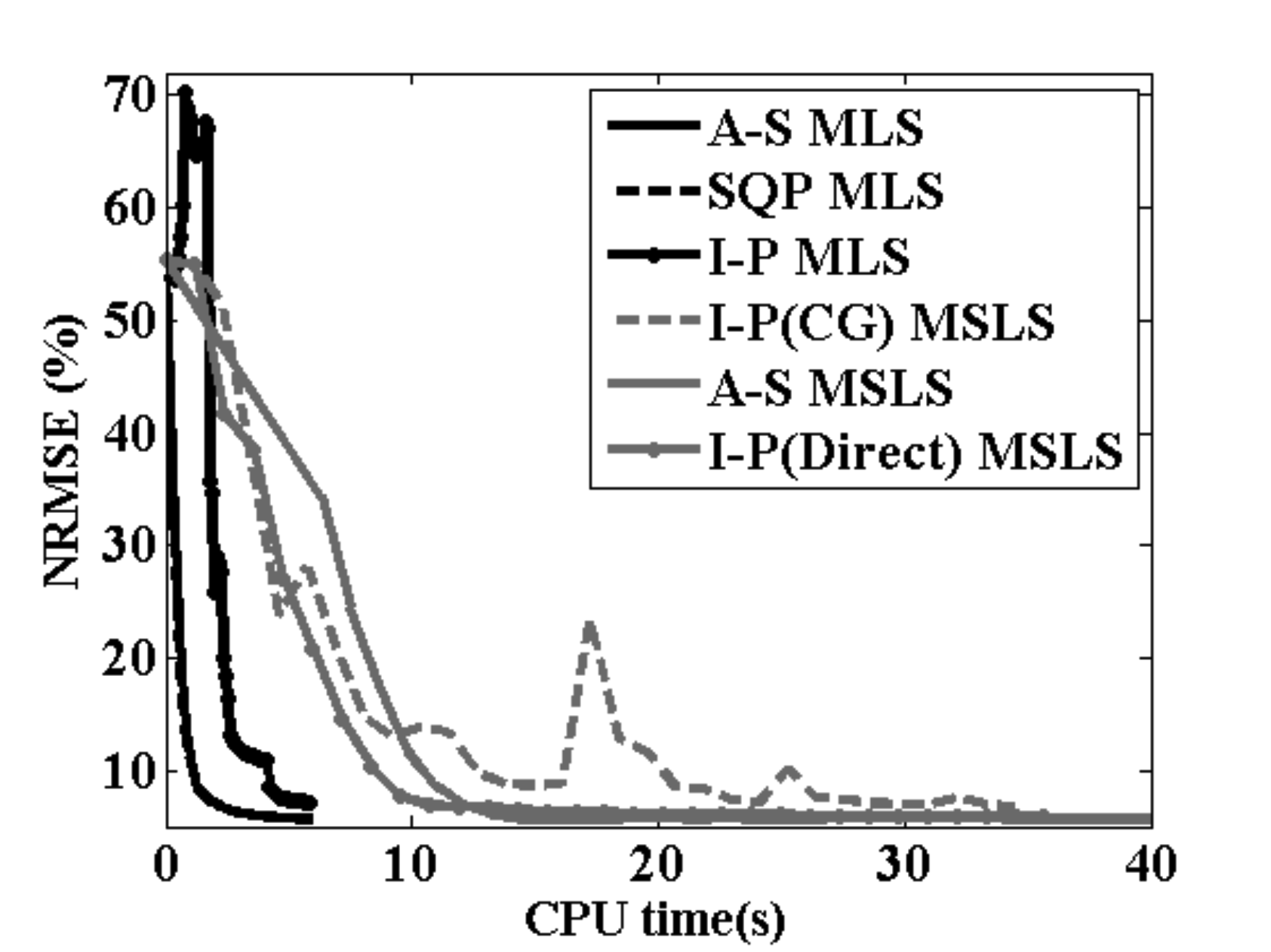}}
  \subfigure[High flip angle regime]{\label{fig:costFuncLTA}\includegraphics[width=4.5cm]{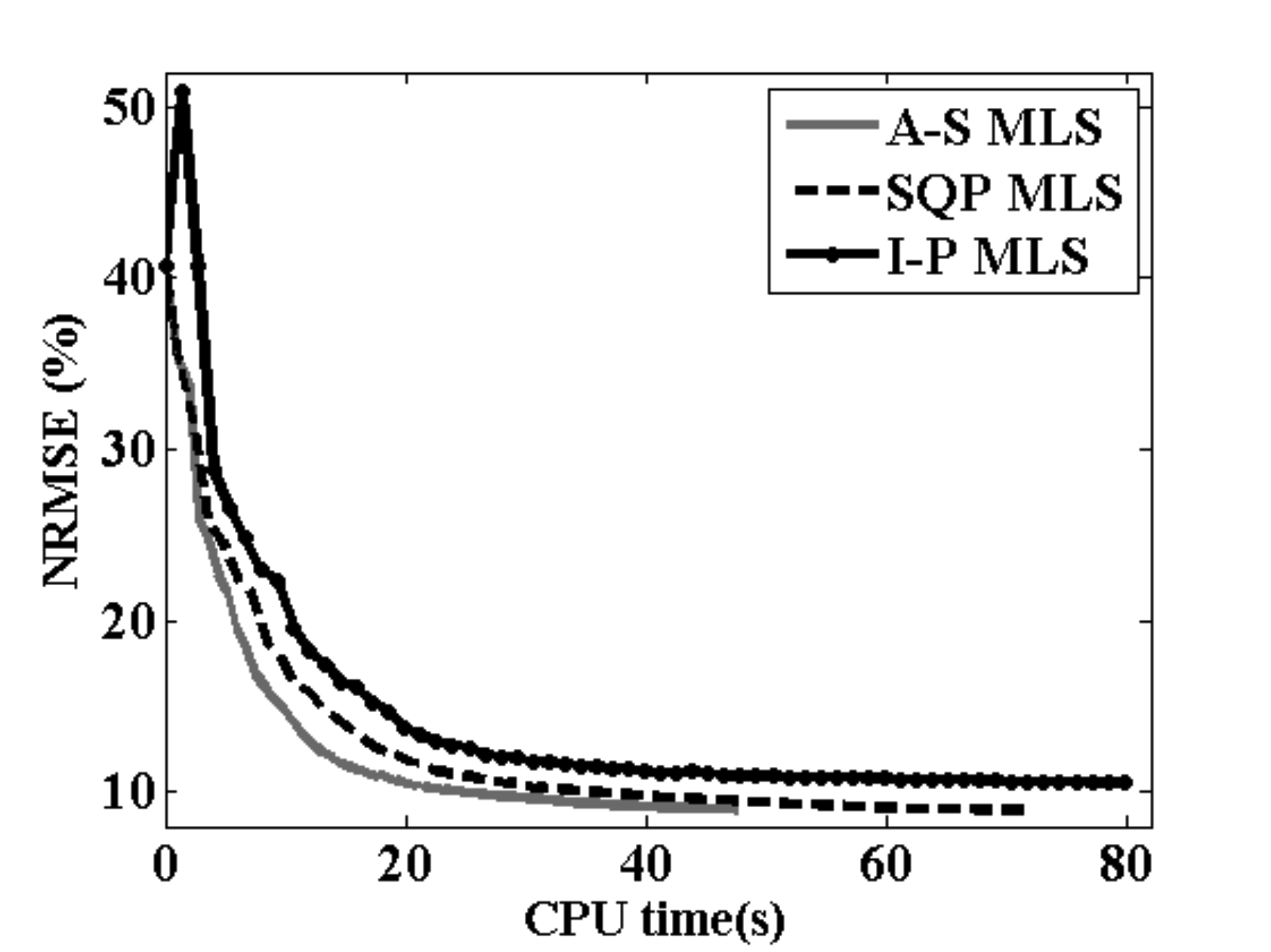}}
  }
\caption{NRMSE versus CPU time for the different algorithms in the small \subref{fig:costFuncSTA} and large \subref{fig:costFuncLTA} FA regimes. The plots for the MSLS variant in the large FA regime are not included due to the time-consuming calculation of the Hessian of the Lagrangian, resulting in a much longer execution time ($\sim 20$ min). In the small FA regime the plots corresponding to the SQP and A-S algorithms are almost indistinguishable. Here, for all algorithms, the initial starting point was chosen based on two criteria: feasibility with respect to all constraints and best NRMSE obtained among all numerical trials performed in this study. This best NRMSE was obtained with a random input vector, thus yielding initially a high cost function.} 
\label{fig:CostFunc}
\end{figure}

The best NRMSEs found therefore are 5.65 and 8.72 \% for the $30^\circ$ and $180^\circ$ target FAs respectively. With a tolerance of 0.3 \% for the NRMSE, when starting from a random initial guess the probabilities to converge to these results were for the different algorithms respectively 84, 84, 0, 86, 88, and 69 \% (same order as in table \ref{tab1}) for the small FA target, indicating a high robustness of the algorithms with a cold start input, except for the I-P implementation of Matlab. The numbers are even more encouraging when the initializations are performed via the G-S algorithm, indicating a very high robustness for these algorithms, especially in the A-S and SQP implementations, when a warm start is provided. These numbers are 100, 100, 0, 90, 91, and 60 \%.  The A-S and SQP methods thus are seen to be very robust with a cross-correlation between the final output solutions varying between 0.98 and 1. As far as the large FA target is concerned, the results indicate, perhaps not surprisingly, a higher sensitivity with respect to the initial starting point when setting a tolerance of 2 \% on the NRMSE, this time the I-P method being the most robust (see Fig. \ref{fig:LTAbox}). With that tolerance, the probabilities were 70, 75, 84, 36, 66 and 28 \% for the random initializations; and 62, 61, 83, 68, 86 and 30 \% for the G-S initialization. The I-P (MLS) approach as a result seems more robust for the large flip angle case but yields slightly worse NRMSE on average than the A-S and SQP methods (see Fig. \ref{fig:LTAbox}), and can take almost twice longer to execute (see Fig. \ref{fig:costFuncLTA}). Moreover, for the G-S initializations it is worth pointing that as soon as the Tikhonov parameter exceeded the value of 1000, all MLS methods converged towards their respective minimum with 100 \% probability. The A-S or SQP method combined with a G-S initialization and a strong power regularization hence still appears the method of choice in the large FA regime.
Despite the differences between the original MLS problem and the MSLS reformulation, the numerical experiments performed indicate (but do not prove) that the exact knowledge of the Hessian is not required and that much faster, but approximate, methods such as the BFGS-update method return equally good results. To illustrate convergence speed, we calculated the NRMSE versus CPU time for each algorithm and by selecting the initial starting point which both verifies the constraints and leads to the best NRMSE achieved in all runs.

\begin{figure*}[htb]
 \centering
 \centerline{
  \subfigure[Low flip angle regime]{\label{fig:FAmapSTA}\includegraphics[width=8cm]{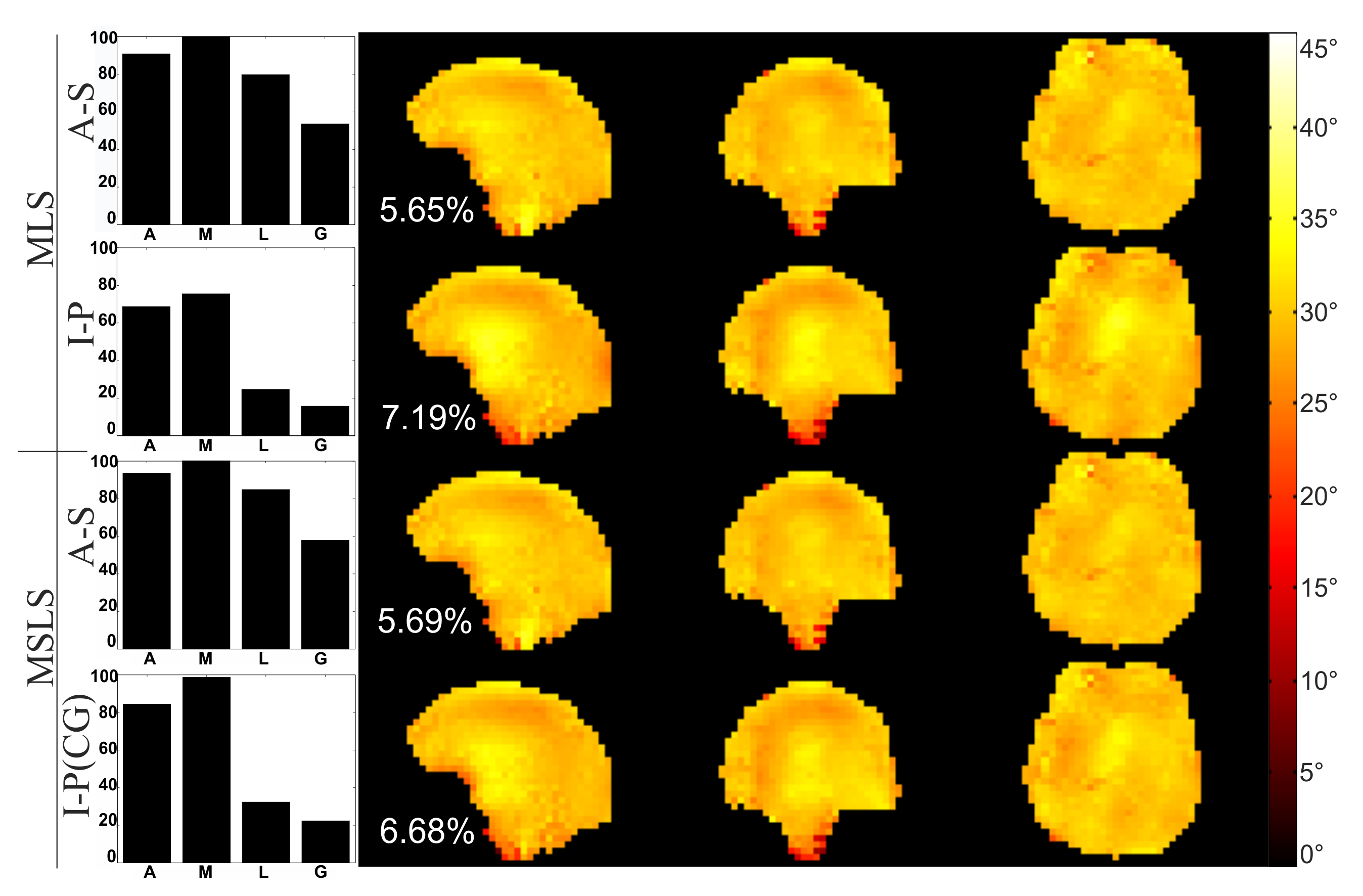}}
  \subfigure[High flip angle regime]{\label{fig:FAmapLTA}\includegraphics[width=8cm]{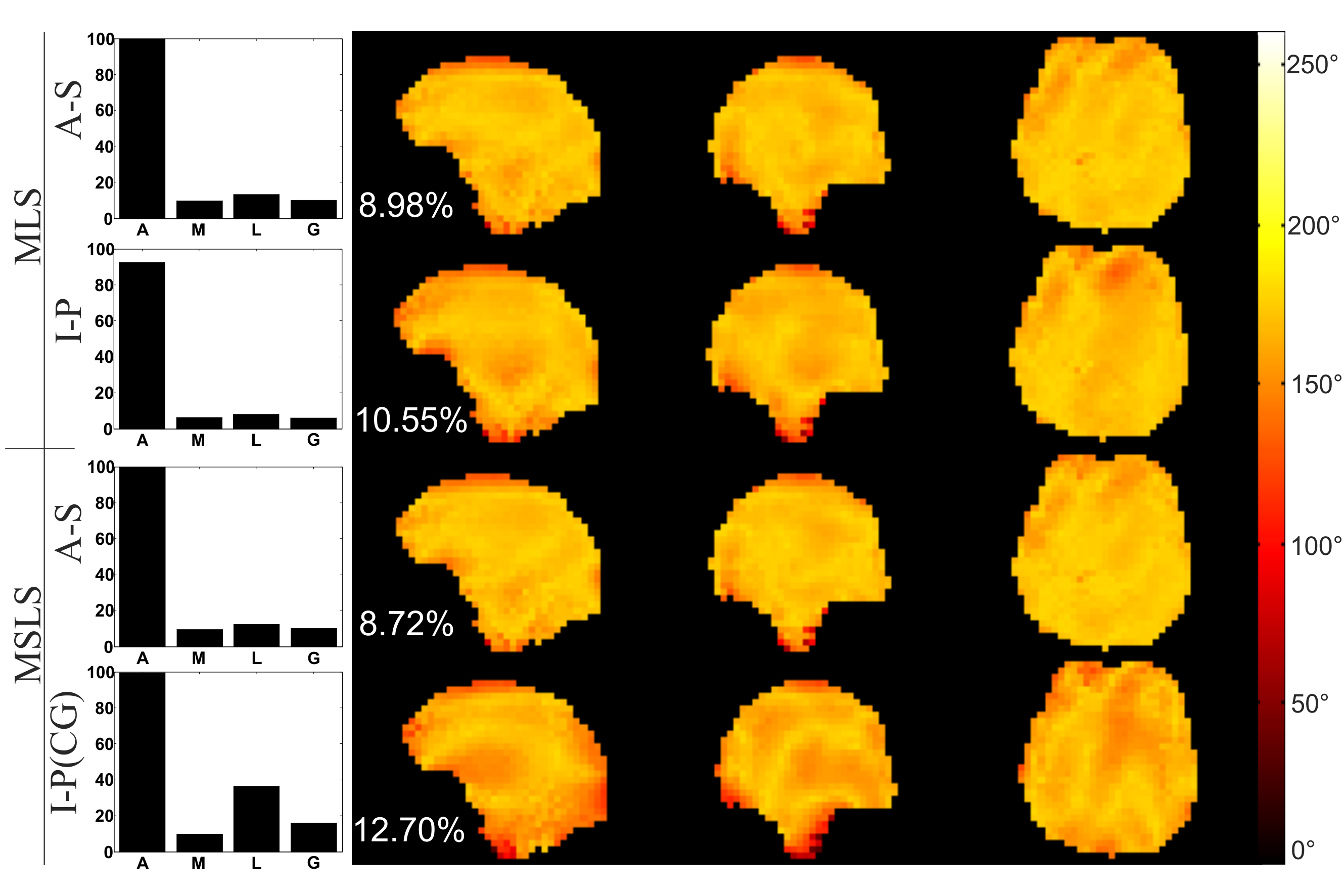}}
  }
\caption{Flip-angle maps obtained for the small \subref{fig:FAmapSTA} and the large tip angle regimes \subref{fig:FAmapLTA}. The bar graphs on the left side of the maps indicate in a percentile way how far from their limits the constraints are ("A" is for amplitude, "M" for the maximum average power among the different channels, "L" for local SAR and "G" for global SAR). The numbers in the left of the FA maps correspond to the NRMSE.}
\label{fig:FAmap}
\end{figure*}

While it seemed not critical for the A-S methods to start with a feasible point to achieve a very good result, the theory behind I-P methods assumes such a starting point. Results are indicated in Figure \ref{fig:CostFunc} both for the small and large FA problems. The corresponding FA maps over the 3 dimensional brain are calculated via a Bloch simulation and are provided in Figure \ref{fig:FAmap}, also with the relative saturation of the different constraints indicated in bar graphs (i.e. how far the returned values are from their respective limits). In our case, it seemed that average power and peak amplitude were the most critical constraints for the small FA design, whereas the most critical constraint in the large FA regime was by far the amplitude. 

\begin{table*}
\caption{Peak power (A) (without units), maximum average power (P) (W), maximum 10-g SAR (L) (W/kg) and global SAR (G) (W/kg) corresponding to Table \ref{tab1}  before and after each algorithm. The initializations here correspond to the ones leading to the lowest NRMSE for each algorithm.}
\centering
\begin{tabular}{cc|c|c|c|c|c|c|c|c|c|}

  \cline{3-10}
  & \multicolumn{1}{c}{}   & \multicolumn{4}{|c|}{After initialization}        & \multicolumn{4}{c|}{After algorithm}   \\
  \cline{3-10}\\[-0.8em]
  \cline{3-10}
   & & A & P & L & G  & A & P &  L & G \\
  \hline
  \multicolumn{1}{ |c| }{\multirow{6}{*}{\rotatebox[origin=c]{90}{SFA}}} 
   & A-S (MLS) & 0.81 & 10.17 & 5.15   & 0.85     &  0.94        & 10.00   & 6.95 & 1.55 \\
   \multicolumn{1}{ |c| }{} & SQP (MLS) & 0.81  & 10.29          & 5.19 &  0.85 &  0.94      & 10.00     & 6.97 & 1.55 \\ 
   \multicolumn{1}{ |c| }{} & I-P Direct \& CG (MLS) & 0.57   &  5.51  & 2.60  &  0.49  &  0.66    & 7.44   & 2.46 & 0.47\\ 
   \multicolumn{1}{ |c| }{} & A-S (MSLS) &   0.37  &  2.81  & 0.95 &  0.22  &  0.94    &  10.00  & 7.67 & 1.73 \\
   \multicolumn{1}{ |c| }{} & I-P Direct (MSLS) & 0.50    &  4.34  & 1.71 &  0.38  &  0.93    &  10.00  & 6.65 & 1.60 \\
   \multicolumn{1}{ |c| }{} & I-P CG (MSLS) &  0.84   &  11.82  & 5.78 &  0.93  &  0.93    &  10.00  & 7.05 & 1.66 \\ 
  \hline
 \multicolumn{1}{ |c| }{\multirow{6}{*}{\rotatebox[origin=c]{90}{LFA}}} 
   & A-S (MLS) & 0.72 & 0.06 & 0.10 & 0.03 &  1.00        & 0.19   & 0.42 & 0.10\\
   \multicolumn{1}{ |c| }{} & SQP (MLS) & 0.83 & 0.07 & 0.12& 0.03 &  1.00      & 0.19     & 0.42 & 0.10\\ 
   \multicolumn{1}{ |c| }{} & I-P Direct \& CG (MLS) & 0.81 & 0.07 & 0.11 & 0.03 &  0.93    & 0.13   & 0.26 & 0.06\\ 
   \multicolumn{1}{ |c| }{} & A-S (MSLS) &  0.50   &  0.03  & 0.05 &  0.01  &  1.00    &  0.19  & 0.38 & 0.10\\
   \multicolumn{1}{ |c| }{} & I-P Direct (MSLS) & 0.65    &  0.05  & 0.08 &  0.02  &  1.00    &  0.18  & 0.38 & 0.10\\
   \multicolumn{1}{ |c| }{} & I-P CG (MSLS) &  0.82   &  0.07  & 0.12 &  0.03  &  1.00    &  0.18  & 0.35 & 0.09 \\ 
  \hline  

\end{tabular}
\end{table*}

\section{Discussion}

A practical evaluation of optimization algorithms is complicated by details of implementation, heuristics and algorithmic options. In this paper, it is worth stressing that the numerical experiments we have carried out made use of different algorithms provided by different solvers (Matlab, Knitro) with their default options. Furthermore we found that variants of the I-P methods (CG and direct) could behave very differently for the MSLS problem. In those same methods, the evolution of the barrier parameter $\mu$ with respect to the number of iterations likewise can have a great impact on the final result \cite{guerin2013local}. Therefore it is not excluded that slightly different implementations of the same class of algorithms could lead to better results. Our goal here however was to provide a readily implementable solution. Surprisingly, it was also found that the BFGS approximation of the Hessian of the Lagrangian led to equally good results as the ones returned when the Hessian was fully calculated, however for the MSLS variant. This is a useful result which allows to significantly speed up the implementation at no performance expense.   

Among all algorithms and initializations we have tested, our recommendation depends on the tolerance one may have on the satisfaction of the constraints. Both A-S and SQP implementations are robust with respect to the initialization in the small FA regime, as long as the input is plausible (generated for instance quickly by the G-S algorithm). In the large FA regime, the best performance was obtained with the A-S method but this time a smaller robustness with respect to the input state was observed. Initializations performed with the G-S method and a high regularization parameter returned the best result in a reliable manner. Whereas A-S is a bit faster than SQP, the constraints may be slightly violated at the end (by a few \% at the most). Although this can be acceptable for instance for the SAR or average power constraints, this solution would not be accepted by the MRI scanner if the peak power limit was exceeded. This however can be checked a posteriori and corrected by truncating or renormalizing the waveforms, provided the performance is not too much affected. The SQP algorithm on the other hand should return a solution leading to a strict nonviolation of the constraints. For the 3D \kT-points tests we have studied, the gain in computation time for the A-S method versus SQP seemed certainly worth the possible post-correction on the waveforms, which would take negligible time. Despite the good results obtained in \cite{guerin2013local}, our results also seem to indicate that it is more efficient to tackle the MLS problem under constraints directly rather than looping over constrained least-squares problems where the phase of the target FA is updated at each iteration \cite{guerin2013local}. 

For the small FA pulse, the best performance we could find was 5.65 \%, which is comparable to our previous results \cite{cloos2012kt,cloos2012parallel} except this time no tuning of parameters was required to enforce all constraints. For the inversion pulse, the best performance obtained was 8.72 \% using the A-S algorithm, the MSLS variant, and a 3.5 ms \kT-points pulse, but in  almost 20 minutes. A slightly higher NRMSE of 8.90 \% on the other hand could be obtained in 47 seconds still using the A-S technique on the MLS problem. This is worse than the 6 \% inhomogeneity we had obtained using an optimal control approach, a pulse duration of 5.9 ms and in around 3 minutes \cite{cloos2012parallel}. Considering however the much smaller number of degrees of freedom we used here (112 versus $94 \ 000$), the substantially shorter pulse duration and the fact that the SAR constraints this time were directly incorporated, this result is quite remarkable and is likely due to the use of second order methods compared to gradient descent approaches \cite{cloos2012parallel}. Furthermore, note that 8.90 \% inhomogeneity is significantly better than what is typically measured at 3 Tesla using a birdcage coil ( $\sim 12$ \%) \cite{boulant2008strongly}. This performance and efficiency to design such pulses thus allows to mitigate the RF inhomogeneity problem at UHF in standard $T_1$-weighted imaging sequences such as the MPRAGE in very reasonable time and thus in routine.  
             
\section{Conclusion}

In this paper we have investigated several initializations and nonlinear programming methods to solve the MLS problem in RF pulse design utilizing parallel transmission at UHF, under strict SAR and hardware constraints, both in the small (linear) and large (nonlinear) FA regimes. Our final recommendation in both cases is the use of an A-S (or SQP) method combined with a G-S initialization with strong power regularization to yield the lowest NRMSE in the most efficient way. Moreover, the combination of these techniques leads to an execution time that is sufficiently small for the approach to be implemented in routine, although the use of parallel computing devices (GPUs in our case) seems at this point a necessity for the design of large FA pulses.

\bibliographystyle{IEEEtran}
\bibliography{RF_Pulse_Design_Andres}

\end{document}